\documentclass[twocolumn,aps,prb,superscriptaddress]{revtex4-2}
\usepackage{amsmath,empheq}
\usepackage{amssymb}
\usepackage{mathrsfs}  
\usepackage{amsfonts}
\usepackage{booktabs}
\usepackage{physics}
\usepackage{graphicx} 
\usepackage{subfigure} 

\usepackage{epsfig}
\usepackage{color}
\usepackage{epstopdf}
\usepackage{xspace}
 \usepackage{rotating}
 \usepackage{appendix}
 \usepackage{chemformula}
 \usepackage{bm}
 \usepackage[colorlinks=true,linkcolor=blue,citecolor=blue,filecolor=blue,urlcolor=blue]{hyperref}
 \usepackage[export]{adjustbox}

 \definecolor{bondiblue}{rgb}{0.0, 0.58, 0.71}

\begin{document}
 \title{Bad topological semimetals in layered honeycomb compounds}
\author{Manuel Fern\'andez L\'opez}
\affiliation{Departamento de F\'isica Te\'orica de la Materia Condensada, Condensed Matter Physics Center (IFIMAC) and
Instituto Nicol\'as Cabrera, Universidad Aut\'onoma de Madrid, Madrid 28049, Spain}
\author{Jaime Merino}
\affiliation{Departamento de F\'isica Te\'orica de la Materia Condensada, Condensed Matter Physics Center (IFIMAC) and
Instituto Nicol\'as Cabrera, Universidad Aut\'onoma de Madrid, Madrid 28049, Spain}
\begin{abstract}
The Mott transition in honeycomb compounds with significant spin-orbit coupling is explored. At finite temperatures we identify a novel semimetallic phase located between a topological insulator and a topological Mott insulator. This semimetal is characterized by having no charge gap, no quasiparticles and gapped spin excitations with non-trivial topological properties. While charge conduction is incoherent involving mean-free paths violating the Mott-Ioffe-Regel limit, spin excitations can be transported ballistically along the edges giving rise to a quantized spin conductance. Such bad topological semimetal could be searched for close to the Mott transition in certain twisted bilayers of transition metal dicalchogenides. 
\end{abstract}
 \date{\today}
 \maketitle
\section{Introduction}
The Mott transition in strongly interacting materials with spin-orbit coupling (SOC) is yet poorly understood \cite{Pesin2010}. Electron correlations can strongly distort conventional band-type topological systems leading to novel states of matter \cite{Balents2014}. A topological insulator (TI) under strong Coulomb repulsion can be transformed into a topological Mott insulator \cite{Pesin2010,Rachel2010,Imada2011} (TMI) supporting spin-only topological edge excitations dispersing between bulk incoherent Hubbard bands in contrast to the 'conventional' electronic topological edge states in TI. Strongly correlated states are found around the Mott transition recently induced in certain layered structures of transition metal dicalchogenides (TMD). \cite{Ghiotto2021}

Unconventional topological metallic phases 
can arise at finite temperatures close to
the Mott transition in the presence of SOC.
This is analogous to the bad metallic behavior
observed close to conventional Mott insulators 
such as the cuprates \cite{Kivelson1995,Hussey2004}. At sufficiently strong
Coulomb repulsion and above a crossover temperature scale \cite{Merino2000,Mezio2017},
much smaller than the Fermi temperature, Landau 
quasiparticles disappear
giving way to a metal with incoherent excitations \cite{Kotliar2013,Kokalj2013}. Hence, the important question arises: can a bad metal (or bad semi-metal) with non-trivial topological properties emerge in proximity to the finite temperature Mott insulator transition? For instance, the large resistivities observed in the iridate pyrochlores \cite{Ramirez2019}, Y$_2$Ir$_2$O$_7$, requires theoretical work which goes beyond the conventional band description of a Weyl semimetal.

A convenient platform for exploring the interplay of Coulomb repulsion and SOC is the Kane-Mele-Hubbard (KMH) model. The $T=0$ phase diagram at half-filling, includes semimetallic (SM) and TI phases (for finite SOC)
while, at strong coupling, 
Mott insulators with AF order occur \cite{Rachel2010}. Also a spin-gapped
quantum spin liquid (QSL) interpreted as a TMI 
has been found in proximity to the Mott transition \cite{Hohenadler2011,Zheng2011,Hohenadler2013,Fiete2011,LeHur2012}, $U \sim U_c$, at weak SOC. A temperature-driven transition from a TI to a metal, whose properties remain unknown, occurs at weak SOC. \cite{LeHur2012} Finite-$T$ 
strongly correlated phases are relevant to twisted WSe$_2$ homobilayers \cite{MacDonald2019,Fu2021a,Georges2021} and 60$^0$ twisted  
Moir\'e MoTe$_2$/WSe$_2$ heterobilayers where an intriguing weakly first order topological Mott-to-QAH transition not accompanied by a gap closure occurs \cite{Fu2021b}. 

\begin{figure}[t!]

    \centering
    \includegraphics[width=8.5cm,clip]{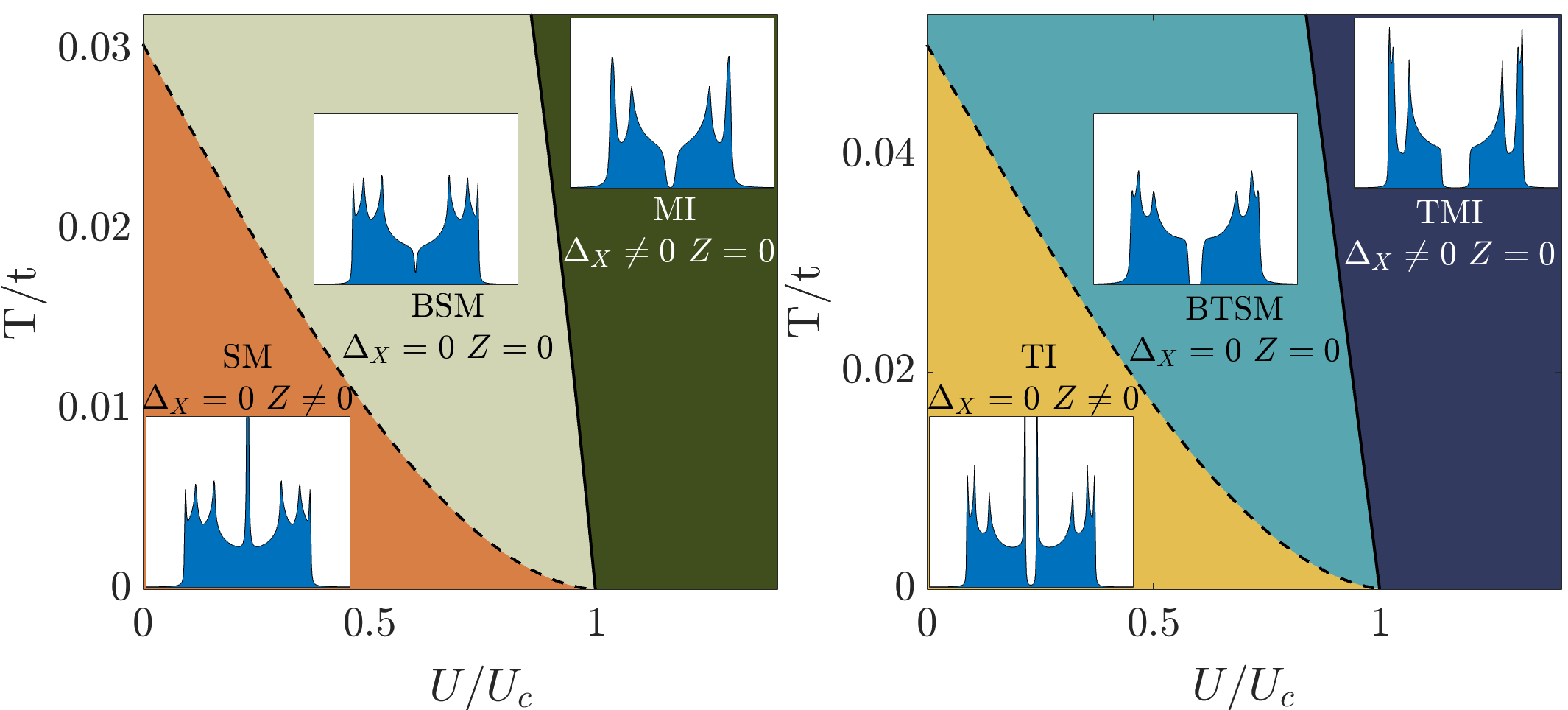}

    \caption{Bad semi-metallic phases in strongly correlated honeycomb lattices. The $T-U$ phase diagram of the Kane-Mele Hubbard model with zero SOC (left panel) is compared with non-zero SOC, $\lambda_{SO}=0.15t$, (right panel) using slave rotor mean-field theory. 
    At $T=0$ a single continuous transition 
    to Mott insulating phases occurs at $U_c$. At finite temperatures and no SOC, a bad semimetal arises as an intermediate phase between a conventional semimetal (SM) and a Mott insulator (MI).
    Such bad semimetallic (BSM)
    phase is characterized by having incoherent ($Z=0$) but gapless ($\Delta_X = 0$) charge-only excitations and ungapped spin-only (spinon) excitations. The bad topological semimetal (BTSM) arising in the presence of SOC, shares the same charge excitation properties as the BSM but has topologically gapped spinons with a $Z_2$ topological invariant, $\nu=1$. The $T=0$ Mott transitions at zero and finite SOC are $U_c=1.68t$ and $U_c=1.82t$, respectively.
    The insets show how the quasiparticle peaks in the electron spectral density, 
    $A_d({\bf k}\sim{\bf k}_F,\omega)$, characteristic of the conventional SM phases are absent in the BSM and BTSM phases. The dashed, $U_{c1}(T)$, and solid, $U_{c2}(T)$, lines represent first and second order transitions, respectively. }
    
    \label{fig:phasediagramT}
\end{figure}

Motivated by the above issues we have obtained the complete finite-$T$ phase diagram of the KMH model. The $T$-$U$ phase diagrams with zero and non-zero SOC are shown in Fig. \ref{fig:phasediagramT} summarizing our main findings. Crucially, we identify a novel bad topological semi-metal (BTSM) arising as an intermediate phase between a topological insulator (TI) and a topological Mott insulator (TMI). In contrast to the bad semimetal (BSM) occurring between the conventional SM and the MI,  
spin-only excitations in the BTSM behave as  
in a $Z_2$ topological insulator \cite{Kane2005b}, while charge-only 
excitations are incoherent in contrast to the
band-like excitations characterizing 
conventional (semi)-metals. Transport properties
of the BTSM include a fractional quantum spin Hall (FQSH) effect due to spinon edge states and bad metallicity {\it i. e.} 
the charge mean-free path, $l$, can become 
smaller than the Mott-Ioffe-Regel (MIR) limit \cite{Pahkira2015}, $l<a$, (with $a$ the lattice parameter) implying the absence of quasiparticles.

\section{Slave-rotor approach to the KMH model}
We consider the Kane-Mele model \cite{Kane2005a,Kane2005b} extended to 
include an onsite repulsive Hubbard interaction:
\begin{align}
    \mathcal{H}=&\sum_{ij\sigma}(t_{ij}e^
    {i\phi_{ij}^\sigma}d_{i\sigma}^\dagger d_{j\sigma}+h.c.)\nonumber\\
    &+\frac{U}{2}\sum_{i}(n_i-1)^2-\mu \sum_{i\sigma} d^\dagger_{i\sigma} d_{i\sigma}
    \label{eq:kmh}
\end{align}
where $t_{ij}$ is the hopping matrix, $t_{ij}=-t$, for nearest-neighbours (NN) and $t_{ij}=e_{ij}\lambda_{so}$ for next-nearest-neighbours (NNN), with $e_{ij}=1(-1)$ for an electron turning left (right) to reach its second neighbor. The spin-dependent complex phase $e^{i\phi_{ij}^\sigma}$ is set to $\phi_{ij}^\sigma=0$ for NN and $\phi_{ij}^\sigma=+(-)\pi/2$ for  $\sigma=\uparrow(\downarrow)$ NNN The electron occupation number operator is defined as $n_i\equiv\sum_{\sigma}d_{i\sigma}^\dagger d_{i\sigma}$

We treat the Kane-Mele-Hubbard (KMH) hamiltonian (\ref{eq:kmh}) within slave-rotor mean-field theory \cite{Florens2002,Florens2004} (SRMFT). The annihilation (creation) electron operator is splitted into a neutral fermion field carrying only the spin (spinon), $f_{i\sigma}$, and a spinless bosonic field carrying only charge (rotor), $X_i$, $i.e.$, $c_{i\sigma}=X_if_{i\sigma}$ ($c_{i\sigma}^\dagger=X_i^*f_{i\sigma}^\dagger$). Thus, 
the original KMH hamiltonian (\ref{eq:kmh}) is splitted into
a spinon and a rotor hamiltonian 
which, at half-filling ($\mu=0$), read:
\begin{align}
    \mathcal{H}_f&=\sum_{ij\sigma}(Q_{ij}^ft_{ij}e^{i\phi_{ij}} f_{i\sigma}^\dagger f_{j\sigma}+h.c.)\\
    \mathcal{H}_X&=\sum_i(\frac{U}{2} L_i^2+\lambda X_i^*X_i)+\sum_{ij}(Q_{ij}^Xt_{ij}X_i^*X_j+h.c.),
    \label{hamiltonianX}
\end{align}
where $Q_{ij}^f\equiv\langle X_i^*X_j\rangle$ and $Q_{ij}^X\equiv\sum_{\sigma}\langle f_{i\sigma}^\dagger f_{j\sigma} \rangle$ are the spinon and rotor renormalization factors, respectively. $L_i$ ($L_i\equiv n_i-1$) is the orbital angular momentum describing the charge quantum number at site $i$, and $\lambda$ is the Lagrange multiplier associated with the constraint, $\langle X_i^*X_i\rangle=1$.

The rotor and spinon Green's functions read: 
\begin{eqnarray}
G_{X\alpha}^{-1}({\bf k},i\nu_n) &=&\nu_n^2/U+\lambda+\epsilon_{\alpha}^X({\bf k})
\nonumber \\
G_{f\alpha}^{-1}({\bf k},i\omega_n) &=& i\omega_n-\epsilon_{\alpha}^f({\bf k}) 
\end{eqnarray}
where $\epsilon_{\alpha}^X({\bf k})$ ($\epsilon_{\alpha}^f({\bf k})$) are the dispersions
associated with the kinetic energy in $H_X$ ($H_f$), respectively. Note that these dispersions contain renormalization effects since they include the $Q_{ij}^f$ and $Q_{ij}^X$ factors. Performing fermionic, $\omega_n=(2n+1)\pi/\beta$, and bosonic, $\nu_n=2n\pi/\beta$, 
Matsubara sums explicitly one arrives at the set of self-consistent equations: 
\begin{align}
1&=\frac{1}{N_cN}\sum_{\alpha\bf k}\frac{U}{2E^X_{\alpha{\bf k}}}\left[b(E^X_{\alpha{\bf k}})-b(-E^X_{\alpha{\bf k}})\right]\nonumber\\
Q^f_{ij}&=\frac{1}{N}\sum_{\alpha\bf k}e^{-i{\bf k }{\bf \delta}_{ij}}\eta_{i}^{\alpha}({{\bf k }}) \eta_{j}^{\alpha*}({{\bf k }})\frac{U}{2E^X_{\alpha{\bf k}}}\left[b(E^X_{\alpha{\bf k}})-b(-E^X_{\alpha{\bf k}})\right]\nonumber\\
\label{eq:Qf}
Q^X_{ij}&=\frac{1}{N}\sum_{\sigma}\sum_{\alpha\bf k}e^{-i{\bf k }{\bf \delta}_{ij}}\chi_{i\sigma}^{\alpha}({{\bf k }}) \chi_{j\sigma}^{\alpha}({{\bf k }}) f(\epsilon_{\alpha}^f({{\bf k}})),
\end{align}
where $\pm E^X_{\alpha{\bf k}}\equiv\pm \sqrt{U(\lambda+\epsilon_{\alpha}^X({{\bf k}}) )}$ can be seen as the energies of an effective quantum harmonic oscillator described by $G_X$. These equations
allow the determination of $Q^f_{ij}$, $Q^X_{ij}$ and $\lambda$ from which the rotor (spinon) eigenvectors and eigenvalues $\eta_{i}^{\alpha}({\bf k})$ ($\chi_{i\sigma}^{\alpha}({{\bf k}})$) 
and $\epsilon_{\alpha}^X({\bf k})$($\epsilon_{\alpha}^f({\bf k})$) are obtained at each iteration.\\ 
\begin{figure*}[t!]
     \includegraphics[width=17cm,clip]{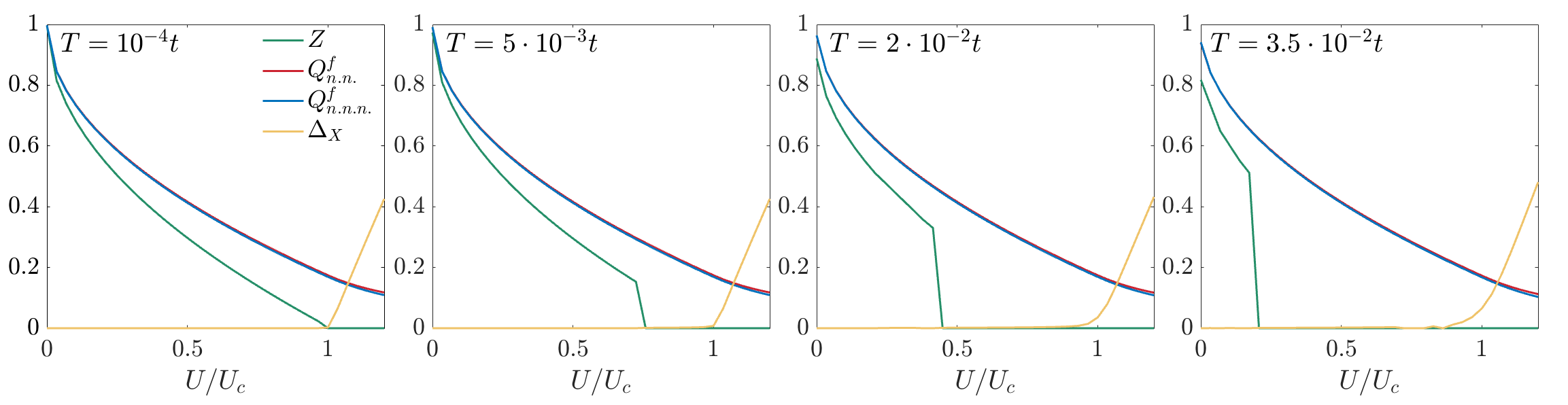}    
    \caption{Dependence of the slave-rotor mean-field parameters with $U$ in the Kane-Mele Hubbard model. The dependence of the quasiparticle weight, $Z$, the spinon renormalization factors, $Q_f$, and the rotor gap, $\Delta_X$ with $U/U_c$ are shown for non-zero SOC, $\lambda_{SO}=0.15t$, and increasing temperature (from left to right). At $T=0$, $Z \rightarrow 0$ and $\Delta_X \neq 0$ continuously 
    indicating a second order TI-to-TMI transition at $U_{c}=U_{c1}(T=0)=U_{c2}(T=0)=1.82t$. In contrast, at finite temperatures, $Z$, drops suddenly at $U_{c1}(T)$ indicating a first order TI-to-BTSM transition while the continuous opening of the rotor gap at $U_{c2}(T)$, signaling the transition to the TMI, remains. In contrast to the drop in $Z$, the spinon renormalization factors, $Q_f$, remains non-zero within the incoherent BTSM and TMI phases, displaying a smooth dependence with $U$. Hence, the spinon bands in both the BTSM and the TMI inherit the topological properties of the TI and due to the bulk-boundary correspondence spinon edge states arise.} 
     \label{fig:Qf_Z_gapX}
 \end{figure*}

In conventional Fermi liquid phases the
Bose-Einstein condensation of the rotors occurs. This is
dealt with numerically by isolating the contribution of the
${\bf k}=0$-mode in the sums of the first two equations 
in (\ref{eq:Qf}):

\begin{align}
1=&Z(T)+\frac{1}{N_cN}\sum_{\alpha{\bf k}\neq 0}\frac{U}{2E^X_{\alpha{\bf k}}}\left[(b(E^X_{\alpha{\bf k}})-b(-E^X_{\alpha{\bf k}})\right] \nonumber \\
Q^f_{ij}=&Z(T)\eta_{i}^{1}(0) \eta_{j}^{1*}({0})+\frac{1}{N}\sum_{\alpha\bf k\neq 0}e^{-i{\bf k }{\bf \delta}_{ij}}\eta_{i}^\alpha({\bf k }) \eta_{j}^{\alpha*}({\bf k })\frac{U}{2E^X_{\alpha{\bf k}}}\nonumber\\
&\times\left[b(E^X_{\alpha{\bf k}})-b(-E^X_{\alpha{\bf k}})\right] ,
 \end{align}
where, $Z(T)$, is the fraction of condensed rotors or the quasiparticle weight which is 
treated as another parameter determined by the self-consistent equations. 
In a one-band model the equation for $Q^f_{ij}$ provides an exact relation between the 
effective mass and the quasiparticle weight \cite{Florens2004}, $Q^f_{ij} = ( \frac{m^*}{m})^{-1}$. At full convergence the rotor gap $\Delta_X$ can be evaluated from: $\Delta_X=\sqrt{U(\lambda+\epsilon_{\alpha}^X({{\bf k}})_{min})}$. On the
other hand $Z$ 
determines whether the metal contains Landau quasiparticles or not.

Since we are interested on thermal effects on the Mott transition
in a two-band model, we include the two rotor bands of the KMH model 
in contrast to $T=0$ works. \cite{Rachel2010} In order to recover 
the correct atomic limit of the KMH model we have performed the rescaling: 
$U \rightarrow U/2$ similar to the one originally introduced 
in the single band Hubbard model \cite{Florens2002,Florens2004}. The TMI predicted by the SRMFT approach is a 
quantum spin liquid (QSL) with a spinon Fermi surface since $t_{ij}^{eff}=t_{ij} Q^f_{ij} \neq 0 $ across the Mott transition. 
In absence of SOC a similar transition to a Mott insulator is captured by a single-orbital Hubbard model on a decorated honeycomb lattice treated by means of slave-bosons theory in the honeycomb-like regime. \cite{Powell2021A,Powell2021B} 
Hence, the electron effective mass does not 
diverge at the Mott transition but $m^*/m \sim U/t$ deep in the Mott insulating phase. \cite{Florens2002,Florens2004} This is in contrast to the Mott transition in infinite
dimensions in which the effective mass diverges \cite{Kotliar1996}, $m^*/m=1/Z \rightarrow \infty$ at $U_c$.

Hence, the correlated phases inherit the topological properties present in the bare Kane-Mele model for $\lambda_{SO} \neq 0$, {\it i. e.} the $Z_2$ topological bulk invariant, $\nu=1$. \cite{Kane2005b,Fu2007} 
Since the SRMFT approach does not include rotor gauge fluctuations
which are known to destroy the TMI in pure two-dimensional systems, 
our results are, strictly speaking, only relevant to layered honeycomb  
structures in which such fluctuations are screened out. \cite{Rachel2018,Young2008} 


Single-electron properties of the model can be characterized by computing the total electron spectral density $A_d({\bf k}, \omega)$, which reads:  
\begin{align}
&A_d({\bf k},\omega)=\nonumber\\
&-\int d{{\bf k '}}\int d\omega' \rho^f_{\bf k}(\omega')\rho_{{\bf k}-{\bf k'}} ^X(\omega-\omega')\left[b(\omega-\omega')+f(-\omega')\right],
\label{eq::resolvedDOS}
\end{align}
where $\rho^{f/X}_{\bf k}(\omega)=-\frac{1}{\pi} \text{Im} G^{f/X}({\bf k}, \omega+i0^+)$ and the total electron density of states (DOS), $\rho_d(\omega)=\int d{\bf k} A_d({\bf k},\omega)$. The above expression 
which is derived in Appendix \ref{spectraldensity} is, in principle, valid at any temperature.

\begin{figure}[t!]
   \centering
\includegraphics[width=8.25cm]{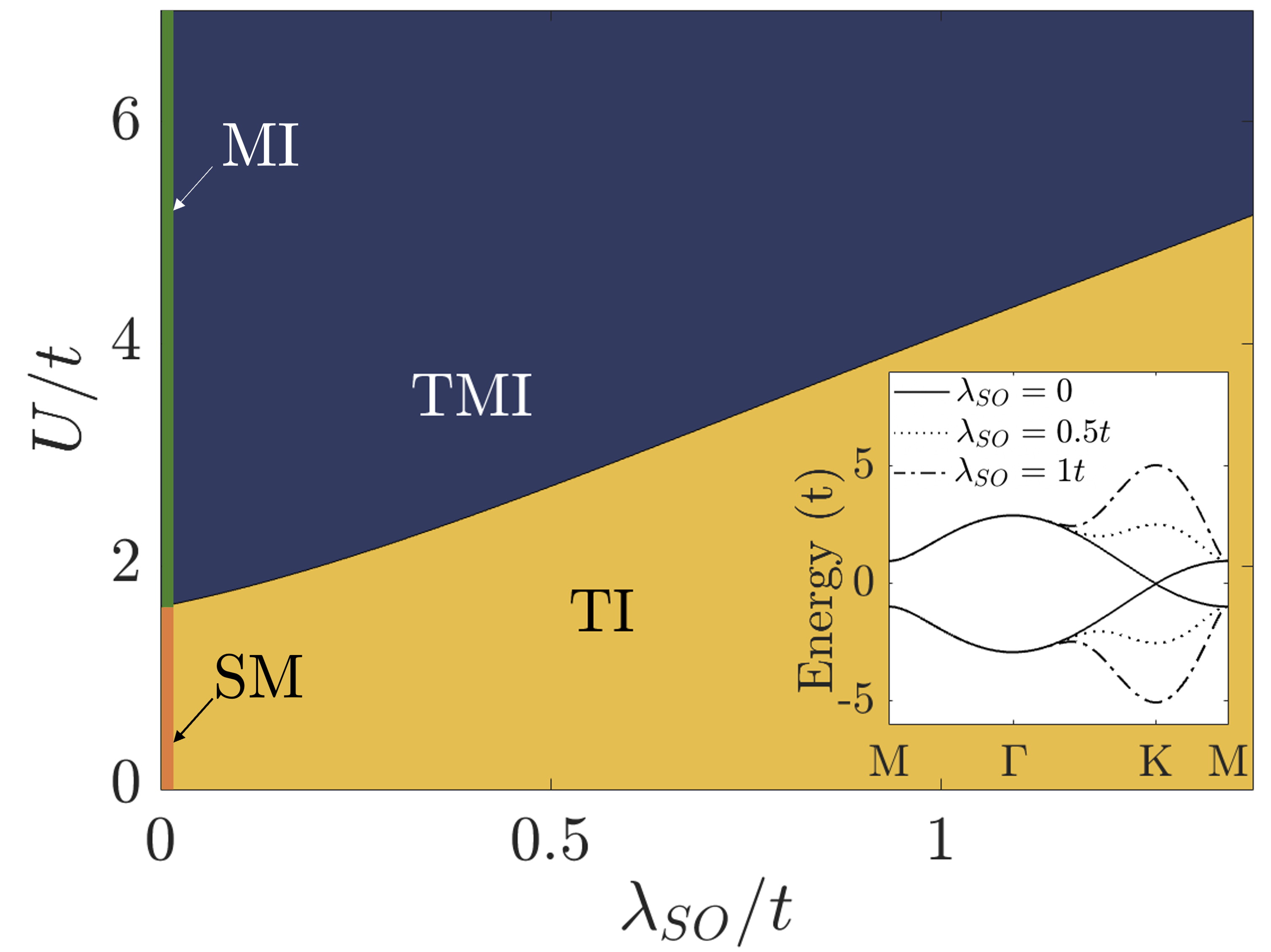}
       \caption{Phase diagram of the KMH model. The critical Hubbard repulsion $U_c$, signaling the $T=0$ TI-to-TMI transition increases with $\lambda_{SO}$. For $\lambda_{SO}=0$ a transition from a SM to a MI occurs. The inset shows the band structure of the KM model for different $\lambda_{SO}=0,0.5t,1t$. }
 \label{fig:phasediagram}
 \end{figure}
\section{Results}
At $T=0$ a continuous Mott transition occurs at
a critical Coulomb repulsion $U_c$.
The quasiparticle weight vanishes, $Z \rightarrow 0$, at $U_c$ 
concomitantly with the charge gap opening, $\Delta_X \neq 0$ within SRMFT \cite{Florens2004}. 
Figure \ref{fig:Qf_Z_gapX} shows that such continuous transition remains when SOC is switched on and the SM to MI transition is converted into a TI to TMI transition.
As previously discussed \cite{Young2008,Rachel2010,Pesin2010}
the TMI is a Mott insulator ($\Delta_X \neq 0$) with a bulk
topological spinon gap and protected spinon edge states leading to 
a FQSH effect. 
For $U>U_c$, within the TMI $\Delta_X$ scales linearly with $U$ as previously 
found in other lattices \cite{lieb}.

The critical $U_c$ increases with SOC as shown in the phase diagram of Fig. \ref{fig:phasediagram}. It varies from $U_c\sim 1.68t$ for $\lambda_{SO}=0$
to $U_c\sim 4.08t$ for $\lambda_{SO}=t$. 
In absence of SOC we recover the $U_c$ found earlier using SRMFT 
techniques. \cite{Lee2005} The $U_c(\lambda_{SO})$ dependence found is due to the bandwidth enhancement expected 
in the KM model when increasing SOC as shown in the inset of Fig. \ref{fig:phasediagram}). 
Such dependence of $U_c$ with SOC is consistent with a previous 
SRMFT study on the KMH model \cite{Rachel2010}.
On the other hand, quantum Monte Carlo (QMC) studies 
of the KMH model predict a much larger critical $U_c=3.89t$ for $\lambda_{SO}=0$. \cite{Hohenadler2011,Zhang2018} This quantitative disagreement with the SRMFT can be attributed to the mean-field nature of the approach in which the electron separation into charge and spin degrees of freedom neglects gauge fluctuations. Including  
such fluctuations at a certain level of approximation 
and imposing the chargon number constraint exactly
could bring the SRMFT $U_c$ closer to the QMC result. 
Such extension of the SRMFT is beyond the scope of the 
present work.

\begin{figure}[t!]
     \centering
     \includegraphics[width=8.5cm,clip]{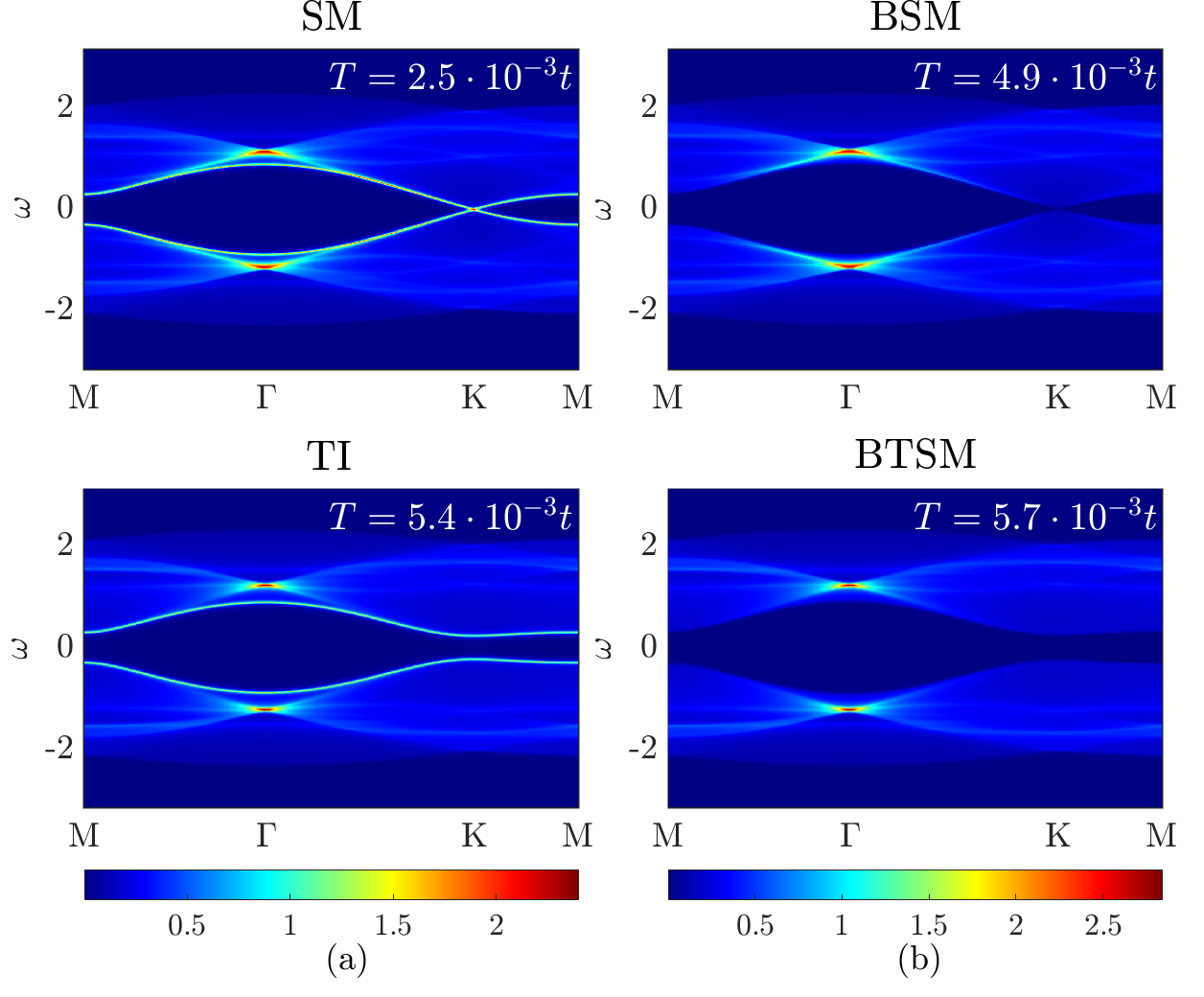}  
     \caption{Destruction of coherent excitations with temperature in the bulk electron spectral density, $A_d({\bf k}, \omega)$. (a) At low temperatures  $T<T_{c1}$ the strongly correlated SM and TI (featured by a $Z_2$ topological bulk invariant $\nu=1$) phases display quasiparticle peaks associated with $Z=0.14$ and $Z=0.15$, respectively. While the SM displays large spectral weight around the Dirac points sharp quasiparticles occur at the gap edges of the TI. 
     $A_d(k,\omega)$ is evaluated in the bulk so surface properties 
     such as topological edge states are not present. 
     (b) At higher temperatures $T>T_{c1}$ only the incoherent ($Z=0$) Hubbard bands formed from the characteristic linear rotor dispersions around the $\Gamma$-point are evident in the BSM and BTSM phases. Coulomb repulsion is fixed to $U/U_c\sim 0.73$ $i.e.$ $U=1.21t,1.34t$ for $\lambda_{SO}=0,0.15t$, respectively. Frequencies are expressed in hopping units.}
     
     \label{fig:resolvedspectraldensity}
\end{figure}

At finite temperatures the Mott transition is characterized by
a two-step process in contrast to $T=0$. 
While at $U_{c1}(T)$ a discontinuous 
transition to a semimetallic ($\Delta_X=0$) but incoherent 
phase ($Z=0$) takes place, at $U_{c2}(T)$
a continuous opening of the charge gap, $\Delta_X \neq 0$  
signalling the Mott insulator occurs. These are first and second order transitions, respectively, as revealed by the 
free energy analysis in Appendix \ref{FreeEnergy}.
As shown in Fig. \ref{fig:Qf_Z_gapX} the 
sudden drop in $Z(T)$ around $U_{c1}(T)$ is in contrast to 
the smooth behavior of spinon, 
$Q_f$, renormalization factors across $U_{c1}(T)$. This leads to a
finite-$T$ region between $U_{c1}(T)$ and $U_{c2}(T)$
with no condensation of chargons, {\it i. e.} $Z=0$, but renormalized
spinon and chargon bands. The $T$-dependence of $U_{c1}(T)$ and $U_{c2}(T)$ leads to a broad asymmetric V-shaped region in the phase diagrams
of Fig. \ref{fig:phasediagramT} in which the incoherent phase is 
stabilized. With no SOC the conventional SM is transformed into an 
incoherent/bad SM (BSM) 
at $U_{c1}(T)$ which is finally converted into a Mott insulator
when $U>U_{c2}(T)$. For non-zero SOC, the conventional TI is transformed
into an incoherent/bad topological SM (BTSM) at $U_{c1}(T)$
becoming a TMI for $U>U_{c2}(T)$. 
Since the spinon renormalization factors, $Q_f$, shown in Fig. \ref{fig:Qf_Z_gapX} remain non-zero, even when no quasiparticles are present, both the BTSM and the TMI have characteristic spinon topological edge states protected by TRS since the spinon bands inherit the topological properties of the TI.

The intermediate incoherent semimetallic phases found are characterized by the absence of well-defined quasiparticles in their single-particle excitation spectrum. The electron spectral density at the Fermi surface, $A_d({\bf k}\sim{\bf k}_F,\omega)$, of the conventional SM and TI phases of Fig. \ref{fig:phasediagramT} 
displays characteristic quasiparticle peaks at the Fermi energy
associated with their underlying band-like coherent excitations, 
as it should. On the other hand incoherent 
Mott-Hubbard bands are evident in the spectra of Mott insulating phases shown in the insets of Fig. \ref{fig:phasediagramT}. In contrast, the excitation spectrum of the intermediate BSM and BTSM phases has no signatures of quasiparticles, only Hubbard-type bands are present. This is striking since these incoherent phases, 
in contrast to actual Mott insulators, are
semi-metallic ${\it i. e.}$ with zero charge gap, $\Delta_X =0$. Intensity plots of $A_d({\bf k},\omega)$ in the ${\bf k}$- $\omega$ plane are shown in Fig. \ref{fig:resolvedspectraldensity}. At fixed $U/U_c \sim 0.73$ as temperature is raised above $T_{c1}$ quasiparticle
dispersions are washed out and only the background of incoherent excitations 
remains. In the SM and TI phases $A_d({\bf k},\omega)$ consists of two different contributions: a linear chargon dispersion centered at the $\Gamma$-point describing Hubbard bands and sharp quasiparticle peaks associated with the spinon band structure containing Dirac band touching points in the SM and a topological 
band gap in the TI.
\begin{figure}[t!]
     \includegraphics[width=8.5cm,clip]{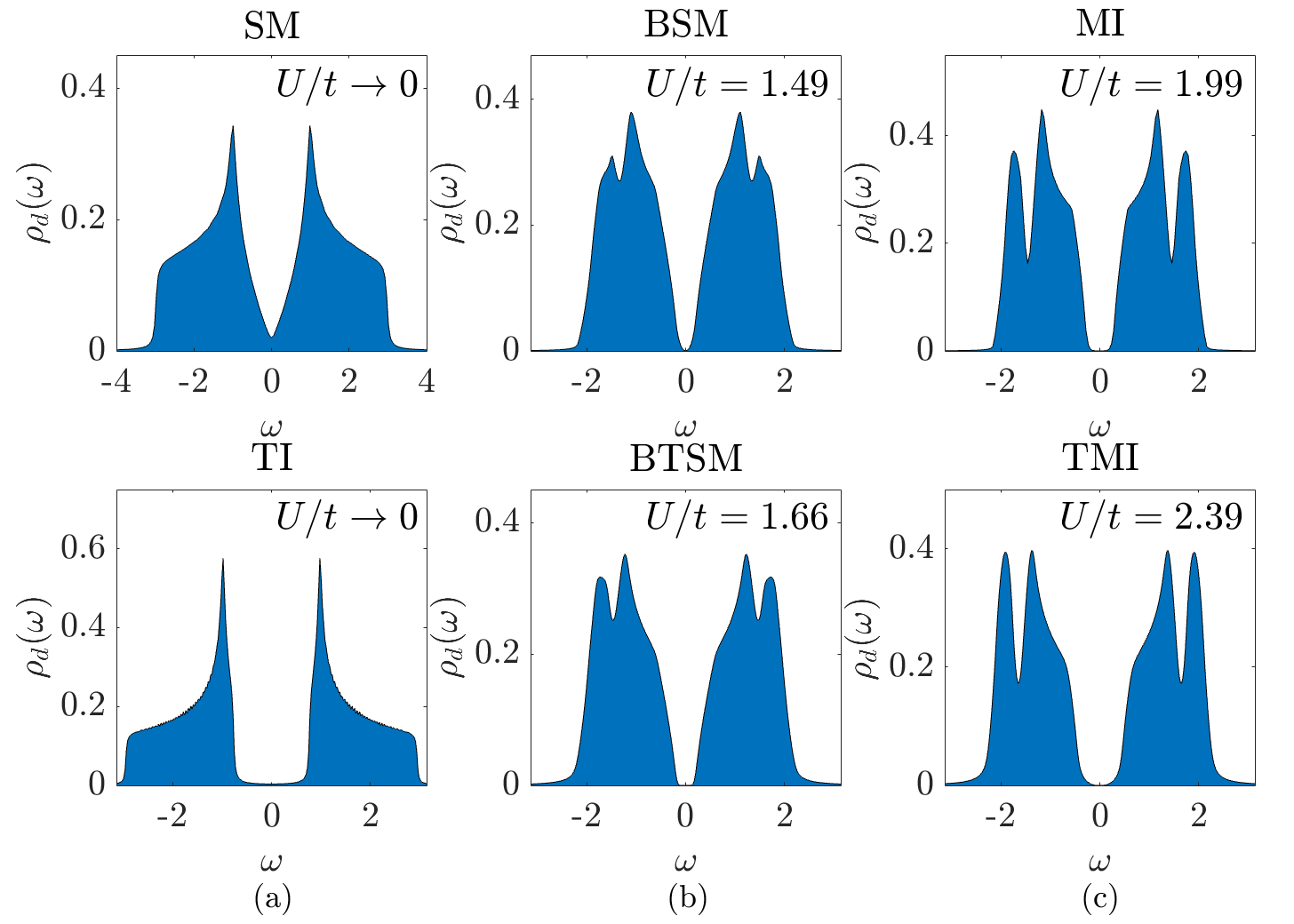}   
     \caption{Incoherent bands in the total electron density of states of the BTSM. The integrated spectral density, $\rho_d(\omega)$, for the Kane-Mele-Hubbard model without (top panels) and with SOC, $\lambda_{SO}=0.15t$ (bottom panels) across the Mott transition are shown at a finite $T\sim 0.01t$. (a) The DOS of the weakly correlated SM 
     (with $Z=0.98$) and TI (with $Z=0.95$) for $U<U_{c1}(T)$ corresponds to the conventional band structure of the honeycomb lattice. (b) For stronger $U_{c1}(T) < U <U_{c2}(T)$ the DOS of the BSM and BTSM have characteristic incoherent ($Z=0$) Hubbard bands which are gapless (BSM) or separated by a topological gap coming from the spinons (BTSM) although the rotors are gapless (spin-charge separation phenomena). (c) At strong coupling,  $U>U_{c2}(T)$, a rotor gap opens up leading to a Mott insulator with well separated Hubbard bands. Since the TMI retains a spinon gap neutral spin edge states are expected leading to a FQSH. Frequencies are expressed in hopping units. }
     \label{fig:spectraldensity}
 \end{figure}
 
Semi-metallic phases with incoherent Hubbard-like bands emerge 
around the finite-$T$ Mott transition.
The dependence on $U$ of the total electron density of states (DOS) $\rho_d(\omega)$ is shown in Fig. \ref{fig:spectraldensity} at finite $T$. In the weakly interacting regime, $U<U_{c1}$, in which condensation of the rotors occurs, $Z \neq 0$,
spinons and rotors recombine into electrons so that the 
DOS in Fig. \ref{fig:spectraldensity} corresponds to that 
of conventional SM and TI phases renormalized by 
interaction effects. At large $U$ the Mott insulator displays
completely formed Hubbard bands separated by the rotor 
gap $\Delta_X\neq 0$, as it should. In the presence of SOC,
the TMI, unlike the MI, has a topological
spinon gap inherited from the TI leading to a 
a fractionalized quantum spin Hall (FQSH) effect. 
Strikingly, the BTSM is characterized by having a 
finite topological spinon gap separating incoherent 
Hubbard-like bands but, in contrast to the TMI, 
it has no chargon gap, $\Delta_X=0$. Hence, the 
BTSM is a spin-charge separated state in which
gapped spinons behave like in a topological 
insulator conferring its topological properties 
while the chargons are incoherent.

\section{Transport properties of the BTSM}
The BTSM will have rather unusual transport properties
since the charge-only excitations are gapless while the spin-only
excitations have a topological gap. 
Unlike the TMI, the BTSM would conduct charge under applied electric
fields since $\Delta_X=0$. However, due to the lack
of coherence, the BTSM would be an incoherent conductor
with resistivities compatible with mean-free paths, $l$,
smaller than the MIR limit, $l<a$ \cite{Kivelson1995, Hussey2004,Merino2000,GunnarssonBook}, (where $a$ is the lattice
constant) and no Drude peak in 
the frequency dependent optical conductivity. \cite{Merino2008} 

Spinons present in the BTSM and the TMI are insensitive to electromagnetic fields since 
they are neutral particles. Hence, one could consider applying 
thermal gradients and measuring the thermal conductivity, $\kappa_{xx}$,
in order to probe spinon edge states. In contrast to the TMI, the metallic 
nature of the BTSM means that both chargons and spinons 
would contribute to $\kappa_{xx}$ making it difficult to isolate
the spinon edge state contribution. 
A similar problem would arise in 
a thermal Hall effect measurement. We have nevertheless computed the thermal Hall conductivity $\kappa_{xy}/T$ due solely to the spinons under a magnetic field, $B$. Although we have only considered the TMI for simplicity, our results are also relevant to the BTSM since it shares the same spinon excitation properties as the TMI. 
\begin{figure}
    \centering
    \includegraphics[width=8.5cm]{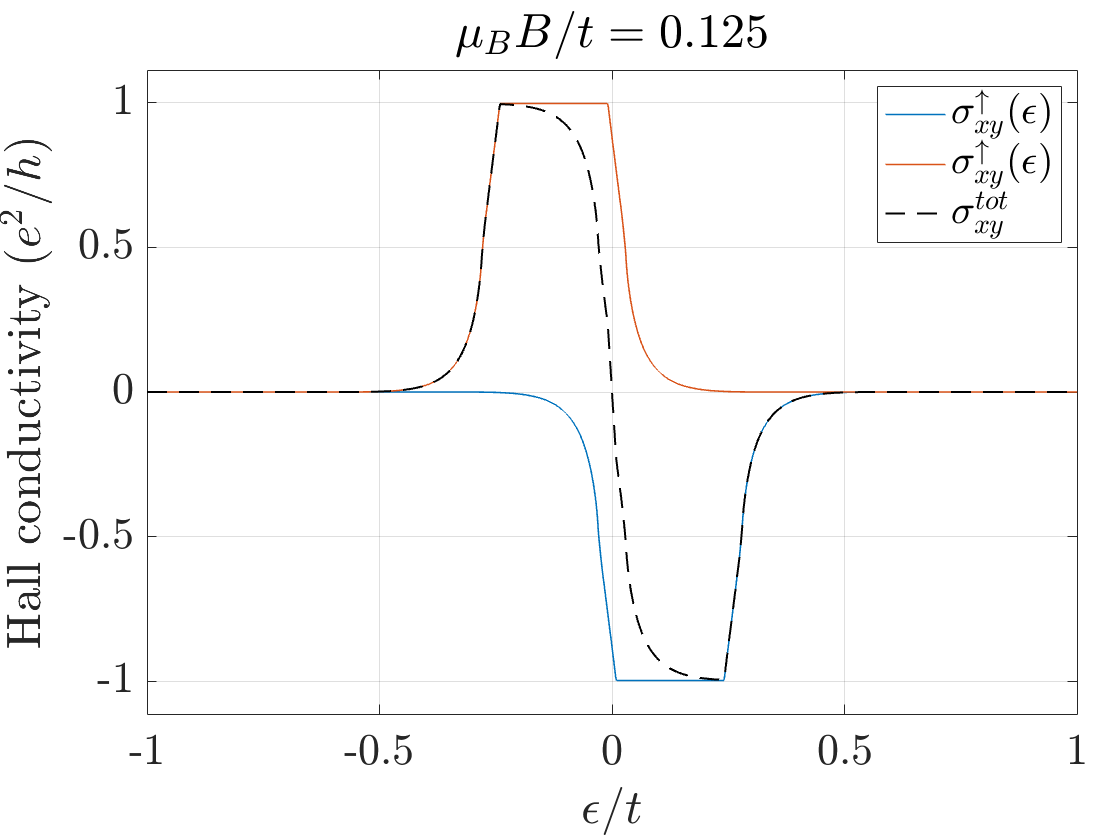}
       \caption{Energy dependence of spin-resolved Hall conductivities, $\sigma_{xy}^{\sigma}$ in the TMI of the KMH model. 
       Within the gap of the TMI 
      $\sigma_{xy}^{\sigma}(\epsilon)= \pm e^2/h$ for $\sigma=\uparrow, \downarrow$ respectively 
       which results from the spin Chern number $\nu_\sigma$ of the occupied band with spin $\sigma$. The spinon energy bands are shifted in opposite directions by $B$  
       leading to a finite total Hall conductivity, $\sigma_{xy}^{tot}(\epsilon)=\sigma_{xy}^\uparrow(\epsilon)+\sigma_{xy}^{\downarrow}(\epsilon)$, which is odd with respect to the Fermi level. This leads to a vanishing thermal Hall 
       conductivity, $\kappa_{xy}/T$ (see \eqref{thermalHall}). Hall conductivities are 
       given here in units of $(e^2/h)$. Calculations are done for $U/U_c= 1.02$, $\lambda_{SO}/t=0.15$ at $T=0$. }
 \label{fig:hallconductivity}
 \end{figure}

We consider only the Zeeman effect of $B$ which only influences 
the spinon part of the SRMFT hamiltonian: ${\Tilde{\mathcal{H}}_f}\rightarrow {{\mathcal{H}}_f+\sigma_z \mu_{B}B}$. The thermal Hall conductivity $\kappa^{2D}_{xy}$ \cite{Shi2011} due to the spinons reads:
\begin{equation}
    \frac{\kappa^{2D}_{xy}}{T}=\frac{\beta^3}{4}\int d\epsilon \frac{(\epsilon-\mu)^2}{cosh^2[\beta(\epsilon-\mu)/2]}\sigma_{xy}^{tot}(\epsilon),
    \label{thermalHall}
\end{equation}
where the Hall conductivity is given by: $\sigma_{xy}^{tot}(\epsilon)=\sigma_{xy}^\uparrow(\epsilon)+\sigma_{xy}^{\downarrow}(\epsilon)$. The spin Hall conductivity depends on the Berry curvature $\Omega_{xy}^{\alpha{\bf k}\sigma}$ through the TKNN formula \cite{TKNN}:
\begin{equation}
    \sigma_{xy}^\sigma(\epsilon)=\sum_{\alpha {\bf k}}\Omega_{xy}^{\alpha\sigma}({\bf k})\theta(\epsilon-\epsilon_{\alpha \sigma}^f({\bf k})).
    \label{eq:hall}
\end{equation}
The Hall conductivities in the TMI are shown in Fig. \ref{fig:hallconductivity} 
for $B \neq 0$. The external magnetic field $B$ leads to a non-zero Hall conductivity $\sigma_{xy}^{tot}(\epsilon)\neq 0$ due to the opposite shift of the spinon bands having opposite spin. However, since $\sigma_{xy}^{tot}(\epsilon)$
is an odd function of the energy with respect to the Fermi energy, the thermal Hall conductivity in Eq. \eqref{thermalHall} vanishes. This is due to the particle-hole symmetry of the spinon-bands expected in the KMH model at half-filling. Thus, in order to have a non-zero thermal Hall conductivity particle-hole symmetry should be broken. This could be achieved  
by electron (hole) doping the model off half-filling. \cite{Lee2019}

From the above discussion it is clear that a neutral spin resolved probe is required to detect the spinon edge states in the BTSM (and also in the TMI). A
possible way could be through spin transport experiments 
in which the BTSM is placed between two spin baths. Under an external spin bias, $V_s=\mu_{\uparrow}-\mu_{\downarrow}$ (where $\mu_\sigma$ denotes the 
chemical potential spinon with spin-$\sigma$) 
applied, say, to the left bath a spin current of magnitude
$I_s=\frac{V_s}{2 \pi}$ \cite{Zekun2021} will flow from left to right 
due solely to spinon edge states. Note that perfect 
quantization of this spin conductivity would only occur 
well below the temperature scale set by the spinon gap, $\Delta_f(U) \approx  Q_f  \Delta_f(U=0)$
which for $U \sim U_c$ leads to $T\ll\Delta_f(U) \sim 0.3t$. 
The spin accumulation in the
left bath can be achieved through the QSH effect
while an inverse QSH effect can be used to detect the spin current on the right lead.
\cite{Sachdev2015,Wang2013} 
 
\section{Conclusions and outlook}
We conclude describing some preliminar evidence of the BTSM in
'exact' numerical treatments
and a discussion of their relevance to actual materials. 
Cluster dynamical mean-field theory calculations \cite{LeHur2012} 
on the KMH model find that the charge gap closes at a low temperature scale, 
$T \sim 0.04t$, above the TI to QSL/SDW Mott transition 
giving a phase diagram consistent with our Fig. 1 (neglecting 
the SDW phase). The spectral density, $A_d(\omega)$, 
of the TI close to the Mott transition as described by DMFT \cite{Kawakami2012}
displays Hubbard and quasiparticle bands below the coherence
scale with a non-zero spin Hall conductivity. Extension to 
temperatures well in the BTSM should find a purely 
incoherent $A_d(\omega)$ consistent with our Fig. \ref{fig:spectraldensity}. Since
the KMH may be realized in 60$^0$ twisted MoTe$_2$/WSe$_2$ heterobilayers \cite{Fu2021b}, 
they are candidates for the BTSM phase. Indeed, resistivities above the MIR limit, $\rho(T) > \hbar a/e^2$ have been 
reported \cite{Georges2021} suggesting bad metallic behavior.  
Spin current injection into these strongly correlated phases  
can probe spinon edge states through measurements of the 
spin conductivity. The recent 
reinterpretation of the pyrochlore iridate, Y$_2$Ir$_2$O$_7$, 
as a bad Weyl semimetal \cite{Ramirez2019} suggests that more topological materials
with bad semimetallic behavior will be discovered in the near future.




 \acknowledgments
We acknowledge financial support from (RTI2018-098452-B-I00) MINECO/FEDER, Uni\'on Europea and the Mar\'ia de Maeztu Program for Units of Excellence in R\&D (Grant No. CEX2018- 000805-M).
\appendix

\section{Local electron spectral density}
\label{spectraldensity}
Single-electron properties are analyzed from the electron spectral density, 
$A_d({\bf k},\omega)$:
\begin{equation}
A_d({\bf k},\omega)=-\frac{1}{\pi}\lim_{\eta \rightarrow 0}\Im{G_d({\bf k},\omega+i\eta)},
\label{eq::spectraldensity}
\end{equation}
where $G_d({\bf k},i\omega_n)$ is the Matsubara electron Green's function.
Within slave-rotor mean-field theory (SRMFT) the imaginary-time Green function of the electron $G_{d}({\bf k},\tau)$ can be expressed as the convolution of the spinon and rotor ${\bf k}$-propagators \cite{Florens2004,Florens2002}:
\begin{align}
    &G_d({\bf k},\tau)=\nonumber\\
    &\frac{1}{\beta}\sum_{nn'}\int d {\bf k}'G_f({\bf k},i\omega_{n'})G_X({\bf k}-{\bf k}',i\nu_n)e^{-i(\omega_{n'}+\nu_n)\tau},
\end{align}
where $\nu_n=2n\pi/\beta$ and $\omega_n=(2n+1)\pi/\beta$ are the bosonic and fermionic frequencies respectively.After carrying out the respective Matsubara sums and
Fourier transforming $G_d({\bf k},\tau)$ to Matsubara frequencies, through \eqref{eq::spectraldensity}, one recovers
the spectral density expression shown in the main text \eqref{eq::resolvedDOS}.
\begin{figure*}[t!]
   \centering
\includegraphics[width=14cm]{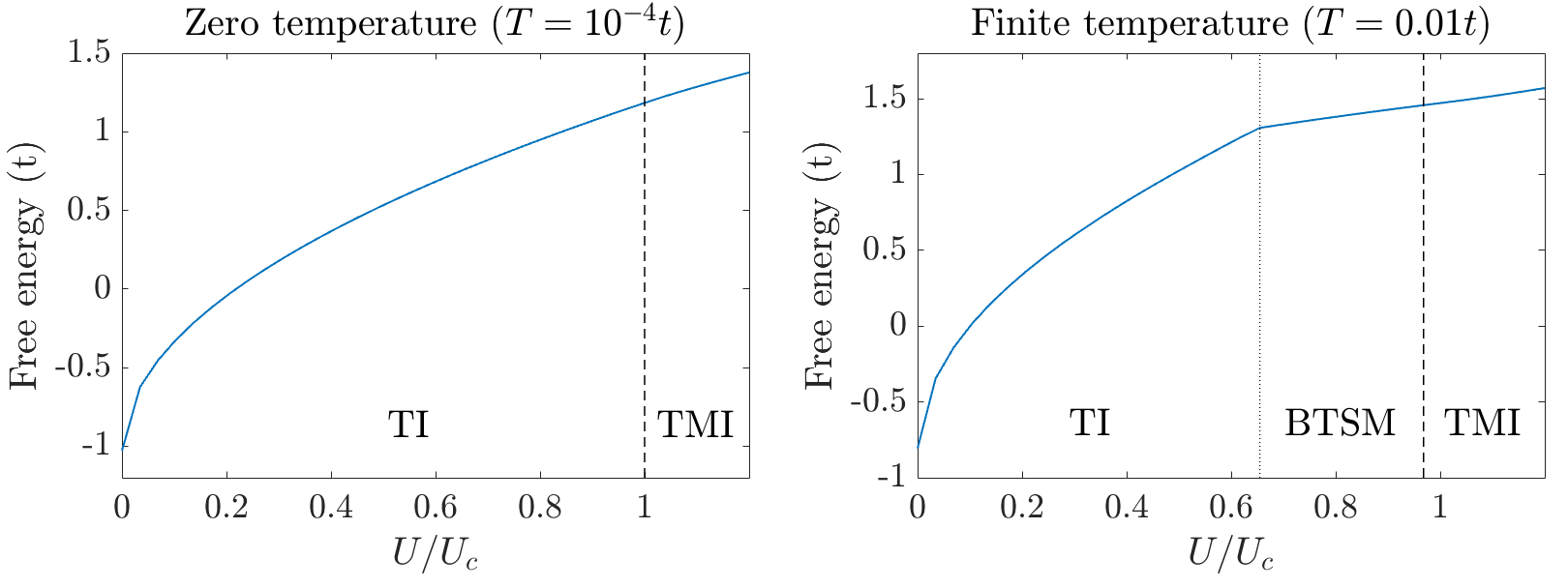}
       \caption{Free energies of the KMH in the presence of SOC ($\lambda_{SO}=0.15t$). (Left panel) At low temperatures ($T/t<10^{-4}$) the free energy behaves smoothly at the TI-to-TMI Mott transition (dashed line) indicating a second order phase transition. (Right panel) At finite temperatures a discontinuity in the free energy derivative occurs at the TI-to-BTSM transition (dotted line) indicating a first order transition.}
 \label{fig:freeenergies}
 \end{figure*}
In phases in which there is a rotor condensed fraction, ${\it i. e.}$ $Z(T) \neq 0$, 
$A_d({\bf k},\omega)$ contains a coherent or quasiparticle contribution:
\begin{align}
    A_d^{coh}({\bf k},\omega)={Z(T)}\delta(\omega-\epsilon^f_{\alpha}({{\bf k}})).
\end{align}
The absence of this contribution, $Z(T)=0$, is a feature of an
incoherent phase. 


\section{Free energy analysis of transitions}
\label{FreeEnergy}
We now elucidate the nature of the transitions described in the main text 
by analyzing the behavior of the free energy of the various phases.
The SRMFT Helmholtz free energy $\mathcal{F}=\mathcal{F}_T+\mathcal{F}_{MF}$ reads:
\begin{align}
&\mathcal{F}_T=-\frac{1}{\beta}\sum_{\textbf{k},\alpha}\left(log[1+e^{-\beta\epsilon^f_{\alpha}(\textbf{k})}] + log[1-e^{-\beta E^X_{\alpha\textbf{k}}}] \right),\\
&\mathcal{F}_{MF}=-\frac{1}{N_c}\sum_{ij}t_{ij}Q^f_{ij}Q^X_{ij}+\lambda,
\end{align}
where spinless rotors $X_i$ and neutral spinons $f_{i\sigma}$ obey Bose-Einstein and Fermi-Dirac statistics respectively. 
In Fig. \ref{fig:freeenergies} we analyze the dependence of the free energy on $U$
at different $T$. At $T=0$ the free energy behaves 
smoothly across $U_c$ where the TI-to-TMI transition occurs. This indicates a second order phase transition. Such single continuous Mott transition becomes a two-step process at larger temperatures since an intermediate BTSM emerges in the region $U_{c1}(T) < U(T) < U_{c2}(T)$ between the 
TI and the TMI. At the TI-to-BTSM transition there is a discontinuity in the free energy derivative due to the crossing of their ground states energies indicating a first order transition. In contrast the BTSM-to-TMI transition is second order since the free energy behaves smoothly across the transition. 
\bibliography{biblio} 
 \end{document}